\documentclass[3p,times,procedia]{elsarticle}
\usepackage{nupha_ecrc}

\volume{00}

\firstpage{1}

\journalname{Nuclear Physics A}

\runauth{Y. Akamatsu, A. Mazeliauskas, D. Teaney}

\jid{nupha}

\jnltitlelogo{Nuclear Physics A}

\usepackage{amssymb}
\usepackage{amsthm}
\usepackage{bm} 

\usepackage[figuresright]{rotating}

\begin{document}

\begin{frontmatter}



\dochead{XXVIth International Conference on Ultrarelativistic Nucleus-Nucleus Collisions\\ (Quark Matter 2017)}

\title{A kinetic regime of hydrodynamic fluctuations and long time tails for a Bjorken expansion}


\author[label1]{Yukinao Akamatsu}
\author[label2,label3]{Aleksas Mazeliauskas}
\author[label2]{Derek Teaney}

\address[label1]{Department of Physics, Osaka University, Toyonaka, Osaka 560-0043, Japan}
\address[label2]{Department of Physics and Astronomy, Stony Brook University, Stony Brook, New York 11794, USA}
\address[label3]{Institut f\"{u}r Theoretische Physik, Universit\"{a}t Heidelberg, 
69120 Heidelberg, Germany}

\begin{abstract}
We develop a set of kinetic equations for hydrodynamic fluctuations which are equivalent to nonlinear hydrodynamics with noise.
The hydro-kinetic equations can be coupled to existing second order hydrodynamic codes to incorporate the physics of these fluctuations, which become dominant near the critical point.
We use the hydro-kinetic equations to calculate the modifications of energy momentum tensor by thermal fluctuations from the earliest moments and at late times in Bjorken expansion.
The solution to the kinetic equations precisely determines the coefficient of the first fractional power ($\propto \tau^{-3/2}$)  in the energy momentum tensor gradient expansion.
Numerically, we find that the contribution to the longitudinal pressure from hydrodynamic fluctuations is comparable to the second order hydrodynamic terms for typical medium parameters used to simulate heavy ion collisions.
\end{abstract}


\end{frontmatter}


\section{Introduction}
\label{sec:intro}
In relativistic heavy-ion collisions, hydrodynamics has been a key concept that makes it possible to understand complicated experimental data in a simple way.
The discovery of collective flow and even-by-event fluctuations of higher flow harmonics in the heavy-ion collisions indicate creation of a strongly interacting quark-gluon plasma whose properties are consistent with a nearly perfect fluid~\cite{Heinz:2013th}.
Currently, a concentrated effort is being made to improve this hydrodynamic paradigm in the heavy-ion collisions~\cite{Yan:2017bgc}.

A qualitatively new ingredient in the hydrodynamic models of heavy-ion collisions is thermal noise~\cite{Kapusta:2011gt}.
Thermal fluctuations is an integral part of a viscous hydrodynamics, as required by the fluctuation-dissipation theorem.
Origin of thermal hydrodynamic fluctuations is the coupling between the slow hydrodynamic degrees of freedom and the fast microscopic degrees of freedom.
Even after integrating out the latter, microscopic degrees of freedom manifest themselves as a stochastic noise to the effective hydrodynamic theory.
So far, most of hydrodynamic simulations for the heavy-ion collisions have neglected the thermal noise, with a few exceptions 
\cite{Kapusta:2011gt,Gavin:2006xd,Young:2014pka,Yan:2015lfa,Murase:2016rhl, Nagai:2016wyx,Gavin:2016hmv}.
Thermal noise in heavy ion collisions may be important, since the number of particles ($\sim$20000) is not so large.
Furthermore, stochastic fluctuations play an essential role near the QCD critical point, and therefore it is important to investigate the effect of thermal noise on the hydrodynamic evolution in heavy-ion collisions.

In this work we focus on the hydrodynamic fluctuations in expanding systems and analyze them in a comoving frame~\cite{Akamatsu:2016llw}.
A specific example  under consideration is the Bjorken expansion, for which the spacetime metric in the comoving frame is $g_{\mu\nu} = {\rm diag}(-1,1,1,\tau^2)$
\footnote{
Another interesting example is a weak shear flow.
Its comoving frame metric is $g_{\mu\nu}={\rm diag}(-1,1+h(t),1+h(t),1-2h(t))$ with $h(t)=h_0 e^{-i\omega t}+{\rm c.c.}$.
Here we can also derive a kinetic theory description and compute the contribution of fluctuations to energy-momentum tensor up to $\mathcal {O} (h_0)$.
We obtain the shear response function $G_R^{\rm shear}(\omega)$, which agrees with the results of \cite{Kovtun:2003vj,Kovtun:2011np} obtained diagrammatically for a static medium.
}.
We find an emergent scale $k_*$ at which the effects of expansion and relaxation of hydrodynamic fluctuations compete with each other.
To describe this regime effectively, we develop a kinetic theory framework for the hydrodynamic fluctuations with wave number $\sim k_*$.
The out-of-equilibrium stochastic fluctuations modifies the energy-momentum tensor through nonlinear terms.
We calculate the slowly decaying nonlinear  contributions to the energy momentum tensor and compare them with second order hydrodynamic corrections.

\section{Kinetic equation for hydrodynamic fluctuations}
\label{sec:kinregm}
In a finite temperature medium hydrodynamic fluctuations are constantly excited  and then propagate and damp according to hydrodynamic equations.
The damping rate for fluctuations with wave number $k$ is given by $\sim \gamma_{\eta}k^2$, where $\gamma_\eta=\eta/(e+p)$ is the momentum diffusion coefficient.
The balance between constant excitation and damping of fluctuations leads to a thermal equilibrium, which is destroyed by the background  expansion.
The Bjorken expansion rate is given by $\sim1/\tau=1/\sqrt{t^2-z^2}$, so hydrodynamic fluctuations with 
\begin{eqnarray}
k\lesssim k_* \equiv \frac{1}{\sqrt{\gamma_{\eta}\tau}}
\end{eqnarray}
relax slowly and are driven out of thermal equilibrium by the expansion.
We call this new emergent scale $k_*$ a kinetic regime. The hydrodynamic gradient expansion is controlled by a small dimensionless parameter $\epsilon\equiv \gamma_{\eta}/c_s^2 \tau \ll 1$.
Using $\epsilon$, the kinetic regime can be expressed as $k_*=\frac{1}{\sqrt{\epsilon}}\frac{1}{c_s\tau}$.
We see that $k_*$ satisfies
\begin{eqnarray}
\frac{1}{c_s\tau} \ll k_* \ll \frac{1}{\epsilon}\frac{1}{c_s\tau}.
\end{eqnarray}
So the kinetic regime is hard compared to the background scale, but is still in the regime controlled by the gradient expansion.
The out-of-equilibrium distribution of the thermal fluctuations is characterized by two point correlation functions.
We will derive evolution equations for the two point correlation functions of these fluctuations in the kinetic regime and analyze how these fluctuations modifies the background flow.

There are four modes of linearized fluctuations $\phi\equiv (c_s\delta e, \vec G)$ in the kinetic regime: left-going and right-going sound modes ($\phi_{\pm}$), and two transverse diffusive modes ($\phi_{T_1,T_2}$).
Here $\delta e$ and $\vec G\equiv (g^x, g^y, \tau g^{\eta})$ are energy and momentum density fluctuations.
Parametrizing the wave vector in the polar coordinates
$\vec K \equiv (k_x,k_y,k_{\eta}/\tau)$, $\hat K\equiv\vec K/|\vec K|=(\sin\theta_K 
\cos\varphi_K, \sin\theta_K\sin\varphi_K,\cos\theta_K)$,
the four modes are given by
\begin{eqnarray}
&&\phi_{\pm}(\tau,\bm k)\equiv \frac{c_s\delta e \pm  \hat K\cdot \vec G}{\sqrt{2}},\\
&&\phi_{T_1}(\tau,\bm k)\equiv -\sin\varphi_K G^x + \cos\varphi_K G^y,\\
&&\phi_{T_2}(\tau,\bm k)\equiv \cos\theta_K\cos\varphi_K G^x + \cos\theta_K\sin\varphi_K G^y -\sin\theta_K G^z.
\end{eqnarray}
Since the kinetic regime is hard, these modes get incoherent easily and do not mix with each other (rotating wave approximation).
The kinetic equations for the two point correlators $N_{AA}(\tau,\bm k)\equiv V^{-1}\langle\phi_{A}(\tau,\bm k)\phi_{A}(\tau,-\bm k)\rangle$ ($A=+,-,T_1,T_2$) are obtained as:
\begin{eqnarray}
&&\partial_\tau N_{\pm\pm}
= -\frac{4}{3}\gamma_\eta K^2
\left[N_{\pm\pm} - \frac{T(e_0+p_0)}{\tau} \right]
-\frac{1}{\tau}\left(2+c_{s}^2+\cos^2\theta_K \right)N_{\pm\pm},\\
\label{eq:kin_N11}
&&\partial_\tau N_{T_1T_1}
= -2\gamma_\eta K^2 
\left[N_{T_1T_1} - \frac{T(e_0+p_0)}{\tau}\right]
-\frac{2}{\tau}N_{T_1T_1},\\
\label{eq:kin_N22}
&&\partial_\tau N_{T_2T_2}
= -2\gamma_\eta K^2
\left[N_{T_2T_2} - \frac{T(e_0+p_0)}{\tau}\right]
-\frac{2}{\tau}\left(1+\sin^2\theta_K\right)N_{T_2T_2}.
\end{eqnarray}
From the kinetic equations, we see that the fluctuations get close to equilibrium at large $|\vec K|/k_*\gg 1$.

\section{Contribution of fluctuations to energy-momentum tensor}
\label{contr}
In a Bjorken expansion, the energy density component $T^{\tau\tau}$ is diluted by the expansion and mechanical work done by the longitudinal pressure $\tau^2 T^{\eta\eta}$, as described by equation of motion
\begin{eqnarray}
\frac{d \langle \tau T^{\tau\tau} \rangle}{d\tau} = -\langle \tau^2T^{\eta\eta} \rangle.
\end{eqnarray}
In the expanding medium the distribution of hydrodynamic fluctuations is modified from that in the equilibrium.
The modified distribution of fluctuations affects the background flow through nonlinear terms of fluctuations, which are given by the constitutive relations
\begin{eqnarray}
\label{eave}
\langle T^{\tau\tau} \rangle = e_0  +  \frac{\langle \vec{G}^2 \rangle}{e +p} , \qquad
\label{pLave}
\langle \tau^2T^{\eta\eta} \rangle = p_0 - \frac{4\eta_0}{3\tau}  + \frac{\langle (G^z)^2 \rangle}{e +p},
\end{eqnarray}
and $\langle G^i G^j\rangle$ is given by a linear combination of two point correlators $N_{AA}$'s integrated in the $\bm k$ space.
For example,
\begin{eqnarray}
\langle (G^z)^2\rangle
=\tau\int\frac{d^3K}{(2\pi)}\left[\frac{N_{++}+N_{--}}{2}\cos^2\theta_K + 
N_{T_2T_2}\sin^2\theta_K\right].
\end{eqnarray}
The asymptotic solution of $N_{AA}$ for large $|\vec K|$ gives divergent contributions to the energy momentum tensor, which can be regularized by a hydrodynamic cutoff scale $\Lambda$.
The energy-momentum tensor should not depend on the cutoff scale $\Lambda$ and its explicit dependence is absorbed by making the background fluid quantities depend on $\Lambda$ (renormalization).
\begin{eqnarray}
p\equiv p_0(\Lambda)+\frac{T\Lambda^3}{6\pi^2}, \ \ \
\eta \equiv \eta_0(\Lambda) + \frac{17\Lambda}{120\pi^2}\frac{T(e_0(\Lambda)+p_0(\Lambda))}{\eta_0(\Lambda)}.
\end{eqnarray}
Here $p$ and $\eta$ correspond to the bare quantities $p_0(\Lambda)$ and $\eta_0(\Lambda)$ in the thermodynamic limit ($\Lambda\to 0$), where all the hydrodynamic fluctuations are integrated out and contribute to the pressure and shear viscosity.
After renormalization, $T^{\eta\eta}$ is cutoff independent and the contribution of thermal fluctuations excited from early time moment to late time is numerically computed as
\begin{eqnarray}
\frac{\langle \tau^2T^{\eta\eta} \rangle}{e+p} = \frac{p}{e+p} - \frac{4\gamma_{\eta}}{3\tau}
+\frac{1.08318}{s(4\pi\gamma_{\eta}\tau)^{3/2}}
+\frac{\lambda_1-\eta\tau_{\pi}}{e+p}\frac{8}{9\tau^2}.
\end{eqnarray}
Here, we include the second-order viscous correction to the ideal Bjorken expansion.
The third term on the right hand side has a fractional power and is called ``long-time tail" because of its power law tail in $\tau$.
Its scaling is simply understood by evaluating the contributions to the pressure from the kinetic regime:
\begin{eqnarray}
T\int_{K\sim k_*} d^3 K\sim Tk_*^3 \sim  \frac{T}{(\gamma_{\eta}\tau)^{3/2}}.
\end{eqnarray}

Substituting typical scales for heavy-ion collisions ($s/T^3\simeq 13.5$ \cite{Borsanyi:2013bia,Bazavov:2014pvz}, $\frac{\lambda_1-\eta\tau_{\pi}}{e+p}\simeq -0.8\gamma_{\eta}^2$ \cite{york:2008rr,Bhattacharyya:2008jc}, and $\tau T\sim 4.5$), the longitudinal pressure is estimated to be
\begin{eqnarray}
\frac{\eta}{s}=\frac{1}{4\pi}&:&
\frac{\langle \tau^2T^{\eta\eta} \rangle}{e+p}=\frac{1}{4}\left[
1 - 0.092\left(\frac{4.5}{\tau T}\right) + 0.034\left(\frac{4.5}{\tau T}\right)^{3/2}
- 0.0008\left(\frac{4.5}{\tau T}\right)^{2}
\right], \\
\frac{\eta}{s}=\frac{2}{4\pi}&:&
\frac{\langle \tau^2T^{\eta\eta} \rangle}{e+p}=\frac{1}{4}\left[
1 - 0.185\left(\frac{4.5}{\tau T}\right) + 0.013\left(\frac{4.5}{\tau T}\right)^{3/2}
- 0.003\left(\frac{4.5}{\tau T}\right)^{2}
\right],
\end{eqnarray}
and we find that the long-time tail contribution is larger than the second order viscous corrections for smaller $\eta/s$.
Therefore, to precisely determine the second order viscous coefficients, one needs to simulate hydrodynamics with thermal noise to correctly evaluate a more important contribution of the long-time tail.

\section{Summary and Outlook}
\label{summar}
In this work we studied hydrodynamic fluctuations in the kinetic regime.
We derived the kinetic equations for the hydrodynamic fluctuations, which describe how the distribution of fluctuations is modified by a Bjorken expanding background.
The fluctuations contribute to the background energy-momentum tensor and contain divergences which needs to be renormalized into the background pressure and viscosity.
The remaining finite contribution has a long-time tail which lasts long as a power law correction $\propto \tau^{-3/2}$ to the background fluid, and makes an essential difference from the hydrodynamics without the thermal noise.
The long-time tail is due to the modified distribution of hydrodynamic fluctuations in the kinetic regime.

The presented formalism provides an alternative way to solve hydrodynamics with noise by coupling the background fluid and fluctuations using the kinetic equations.
We can also develop the kinetic theory for a non-conformal fluid and study how the bulk viscosity is renormalized.
Finally, within the framework of hydro-kinetic we can study the critical fluctuations near the QCD critical point and analyze non-equilibrium Kibble-Zurek scaling behavior in expanding systems.

\subsection*{Acknowledgments}
Y.A.'s work at Stony Brook was supported through a JSPS Postdoctoral Fellowship for Research Abroad. 
A.M.'s and D.T.'s work was supported in part by the U.S. Department of Energy under Contract No. DE-FG02-88ER40388.





\bibliographystyle{elsarticle-num}
\bibliography{qm2017_proc}

\begin{thebibliography}{10}
\expandafter\ifx\csname url\endcsname\relax
  \def\url#1{\texttt{#1}}\fi
\expandafter\ifx\csname urlprefix\endcsname\relax\def\urlprefix{URL }\fi
\expandafter\ifx\csname href\endcsname\relax
  \def\href#1#2{#2} \def\path#1{#1}\fi

\bibitem{Heinz:2013th}
U.~Heinz, R.~Snellings, {Collective flow and viscosity in relativistic
  heavy-ion collisions}, Ann. Rev. Nucl. Part. Sci. 63 (2013) 123--151.
\newblock \href {http://arxiv.org/abs/1301.2826} {\path{arXiv:1301.2826}},
  \href {http://dx.doi.org/10.1146/annurev-nucl-102212-170540}
  {\path{doi:10.1146/annurev-nucl-102212-170540}}.

\bibitem{Yan:2017bgc}
L.~Yan,
  \href{http://inspirehep.net/record/1593948/files/arXiv:1704.06643.pdf}{{Hydrodynamic
  modeling of heavy-ion collisions}}, 2017.
\newblock \href {http://arxiv.org/abs/1704.06643} {\path{arXiv:1704.06643}}.
\newline\urlprefix\url{http://inspirehep.net/record/1593948/files/arXiv:1704.06643.pdf}

\bibitem{Kapusta:2011gt}
J.~I. Kapusta, B.~Muller, M.~Stephanov, {Relativistic Theory of Hydrodynamic
  Fluctuations with Applications to Heavy Ion Collisions}, Phys. Rev. C85
  (2012) 054906.
\newblock \href {http://dx.doi.org/10.1103/PhysRevC.85.054906}
  {\path{doi:10.1103/PhysRevC.85.054906}}.

\bibitem{Gavin:2006xd}
S.~Gavin, M.~Abdel-Aziz, {Measuring Shear Viscosity Using Transverse Momentum
  Correlations in Relativistic Nuclear Collisions}, Phys. Rev. Lett. 97 (2006)
  162302.
\newblock \href {http://dx.doi.org/10.1103/PhysRevLett.97.162302}
  {\path{doi:10.1103/PhysRevLett.97.162302}}.

\bibitem{Young:2014pka}
C.~Young, J.~I. Kapusta, C.~Gale, S.~Jeon, B.~Schenke, {Thermally Fluctuating
  Second-Order Viscous Hydrodynamics and Heavy-Ion Collisions}, Phys. Rev.
  C91~(4) (2015) 044901.
\newblock \href {http://dx.doi.org/10.1103/PhysRevC.91.044901}
  {\path{doi:10.1103/PhysRevC.91.044901}}.

\bibitem{Yan:2015lfa}
L.~Yan, H.~Gr{\"o}nqvist, {Hydrodynamical noise and Gubser flow}, JHEP 03
  (2016) 121.
\newblock \href {http://dx.doi.org/10.1007/JHEP03(2016)121}
  {\path{doi:10.1007/JHEP03(2016)121}}.

\bibitem{Murase:2016rhl}
K.~Murase, T.~Hirano, {Hydrodynamic fluctuations and dissipation in an
  integrated dynamical model}, Nucl. Phys. A956 (2016) 276--279.
\newblock \href {http://dx.doi.org/10.1016/j.nuclphysa.2016.01.011}
  {\path{doi:10.1016/j.nuclphysa.2016.01.011}}.

\bibitem{Nagai:2016wyx}
K.~Nagai, R.~Kurita, K.~Murase, T.~Hirano, {Causal hydrodynamic fluctuation in
  Bjorken expansion}, Nucl. Phys. A956 (2016) 781--784.
\newblock \href {http://dx.doi.org/10.1016/j.nuclphysa.2016.02.007}
  {\path{doi:10.1016/j.nuclphysa.2016.02.007}}.

\bibitem{Gavin:2016hmv}
S.~Gavin, G.~Moschelli, C.~Zin, {Rapidity Correlation Structure in Nuclear
  Collisions}, Phys. Rev. C94~(2) (2016) 024921.
\newblock \href {http://dx.doi.org/10.1103/PhysRevC.94.024921}
  {\path{doi:10.1103/PhysRevC.94.024921}}.

\bibitem{Akamatsu:2016llw}
Y.~Akamatsu, A.~Mazeliauskas, D.~Teaney, {A kinetic regime of hydrodynamic
  fluctuations and long time tails for a Bjorken expansion}, Phys. Rev. C95~(1)
  (2017) 014909.
\newblock \href {http://arxiv.org/abs/1606.07742} {\path{arXiv:1606.07742}},
  \href {http://dx.doi.org/10.1103/PhysRevC.95.014909}
  {\path{doi:10.1103/PhysRevC.95.014909}}.

\bibitem{Kovtun:2003vj}
P.~Kovtun, L.~G. Yaffe, {Hydrodynamic fluctuations, long time tails, and
  supersymmetry}, Phys. Rev. D68 (2003) 025007.
\newblock \href {http://arxiv.org/abs/hep-th/0303010}
  {\path{arXiv:hep-th/0303010}}, \href
  {http://dx.doi.org/10.1103/PhysRevD.68.025007}
  {\path{doi:10.1103/PhysRevD.68.025007}}.

\bibitem{Kovtun:2011np}
P.~Kovtun, G.~D. Moore, P.~Romatschke, {The stickiness of sound: An absolute
  lower limit on viscosity and the breakdown of second order relativistic
  hydrodynamics}, Phys. Rev. D84 (2011) 025006.
\newblock \href {http://arxiv.org/abs/1104.1586} {\path{arXiv:1104.1586}},
  \href {http://dx.doi.org/10.1103/PhysRevD.84.025006}
  {\path{doi:10.1103/PhysRevD.84.025006}}.

\bibitem{Borsanyi:2013bia}
S.~Borsanyi, Z.~Fodor, C.~Hoelbling, S.~D. Katz, S.~Krieg, K.~K. Szabo, {Full
  result for the QCD equation of state with 2+1 flavors}, Phys. Lett. B730
  (2014) 99--104.
\newblock \href {http://dx.doi.org/10.1016/j.physletb.2014.01.007}
  {\path{doi:10.1016/j.physletb.2014.01.007}}.

\bibitem{Bazavov:2014pvz}
A.~Bazavov, et~al., {Equation of state in ( 2+1 )-flavor QCD}, Phys. Rev. D90
  (2014) 094503.
\newblock \href {http://dx.doi.org/10.1103/PhysRevD.90.094503}
  {\path{doi:10.1103/PhysRevD.90.094503}}.

\bibitem{york:2008rr}
M.~A. York, G.~D. Moore, {Second order hydrodynamic coefficients from kinetic
  theory}, Phys. Rev. D79 (2009) 054011.
\newblock \href {http://dx.doi.org/10.1103/PhysRevD.79.054011}
  {\path{doi:10.1103/PhysRevD.79.054011}}.

\bibitem{Bhattacharyya:2008jc}
S.~Bhattacharyya, V.~E. Hubeny, S.~Minwalla, M.~Rangamani, {Nonlinear Fluid
  Dynamics from Gravity}, JHEP 02 (2008) 045.
\newblock \href {http://dx.doi.org/10.1088/1126-6708/2008/02/045}
  {\path{doi:10.1088/1126-6708/2008/02/045}}.

\end{thebibliography}







\end{document}